\documentclass[useAMS,usenatbib]{mn2e}
\usepackage{graphicx}
\usepackage{epsfig}
\usepackage{Times}

\title[Low/Hard State X-ray Outburst of SWIFT J1753.5${-}$0127]
{RXTE observations of the Low/Hard State X-ray Outburst of the new x-ray transient SWIFT J1753.5$-$0127}

\author[M.C. Ramadevi, S. Seetha] 
{M.C. Ramadevi$^{1,2}$\thanks{E-mail:ramadevi@isac.gov.in; seetha@isac.gov.in} and S. Seetha$^1$\\
$^{1}$ISRO Satellite Centre, Airport Road, Vimanapura Post, Bangalore,  
Karnataka, India.\\
$^{2}$Department of Physics, University of Calicut, Calicut, Kerala, 
India.} 

\begin{document}

\date{Submitted to MNRAS}

\pagerange{\pageref{firstpage}--\pageref{lastpage}} \pubyear{2006}

\maketitle

\label{firstpage}

\begin{abstract}
We present the results of the analysis of RXTE (Rossi X-ray Timing Explorer) observations of the new x-ray transient, SWIFT J1753.5-0127 during its outburst in July, 2005.  
The source was caught at the peak of the burst with a flux of 7.19e-09 ergs-s$^{-1}$-cm$^{-2}$ in the 3-25 keV energy range and observed until it decreased by about a factor of 10.
The photon index of the power law component, which is dominant during the entire outburst, decreases from $\sim$1.76 to 1.6.
However, towards the end of the observations the photon index is found to increase, indicating a softening of the spectra.
The presence of an ultrasoft thermal component, during the bright phases of the burst, is clear from the fits to the data.
The temperature associated with this thermal component is 0.4 keV.
We believe that this thermal component could be due to the presence of an accretion disk.
Assuming a distance of 8.5 kpc, $L_{X}/L_{Edd} \simeq 0.05$ at the peak of the burst, for a black hole of mass $10~{\rm M}_{\odot}$.
The source is found to be locked in the low/hard state during the entire outburst and likely falls in the category of the x-ray transients that are observed in the low/hard state throughout the outburst. 
We discuss the physical scenario of the low/hard state outburst for this source.
\end{abstract}


\section{Introduction}
Black Hole Candidates are known to undergo spectral state transitions between the different canonical states they enter during an outburst \citep{Remillard,Homan}.
There is a subclass of x-ray transients, the Low/Hard state X-ray Transients (LHXTs) that undergo outbursts which are entirely in the low/hard state throughout the burst \citep{Brocksopp}.
The accretion process during the low/hard state of the x-ray transients is still not well understood and requires an extensive study, for which this subclass of x-ray transients showing LHXT outbursts, seem promising candidates.

The hard x-ray source SWIFT J1753.5-0127 was first detected by the BAT experiment on Swift satellite at RA (J2000) = 17$^{h}$ 53$^{m}$ 28$^{s}$.3 and Dec (J2000) = -01$^{o}$ 27$^{'}$ 09.3$^{''}$, on June 30, 2005 \citep{Palmer}.
It was also observed by Swift-XRT (X-Ray Telescope) on July 1, 2005 and was found to be extremely bright \citep{Burrow}.
The observations by Swift-UVOT (Ultra-Violet/Optical Telescope) do not indicate any temporal variability on 10-1000s timescale and the spectral fits to the ultra-violet spectra give a lower limit for the temperature of the accretion disk as 116,000 K \citep{Still}.
Optical observations on July 2, 2005, reveal the existence of a bright optical counterpart with R $\sim$15.8, which was not visible on the SDSS \citep{Halpern}.
Spectroscopic studies from the optical observations of the source on July 3, 2005 show a blue continuum with a broad, double-peaked H-alpha emission line with an equivalent width of $\sim$3 Angstroms and FWHM $\sim$2000 km/s \citep{Torresa}.
Simultaneous optical and infrared monitoring of the source done on July 11, 2005, show the presence of an IR point source, at the position of the optical counterpart \citep{Torresb}.
Co-ordinated optical and x-ray observations to study the correlated variability in the two bands show that the high-frequency noise correlates well with the x-ray variations, whereas the low-frequency component is absent in the x-ray data \citep{Hynes}.
Radio observations on July 3, 2005, measured a flux density of 2.1 +/- 0.2 mJy at 1.7 GHz.  Further radio observations on July 4 and 5, 2005, indicate the variability of the source.  The radio source is found to be extended on angular scales not greater than 350 milliarcsec \citep{Fender}.
This could imply the presence of a jet, which is associated with the low/hard state of black hole candidates.
However, this source does not follow the usual radio/X-ray correlation of X-ray binaries in the low/hard state \citep{Bel2006}.
INTEGRAL observations of the source from August 10 to 12, 2005 indicate the spectrum to be typical of that of a BHC in the hard state \citep{Bel, Bel2006}.
The PDS of the RXTE-PCA observations of the source show a 0.6 Hz QPO with a shape typically seen in black hole candidates \citep{Morgan}. 
Simultaneous RXTE and XMM-Newton observations of this source on March 24, 2006, near the quiescent state of this source, by \citet{Millera} show the presence of an accretion disk in the low/hard state of this source.

\section{Observations and Data Analysis}
\label{sec:data}
We have analysed 58 TOO (Target Of Opportunity) observations by RXTE \citep{Bradt}, which span about 146 days of the outburst amounting to about 150 ksec of data.
The observations date from July 6, 2005 to November 28, 2005.
We have analysed the data from the observations of ASM, PCA and the HEXTE instruments on RXTE.
The details of the light curve and the spectral analysis of the outburst are given in the following sections.

\subsection{ASM-PCA Light Curve Analysis}
\label{sec:light-curve}
\begin{figure}
\includegraphics[height=6cm,width=8cm,angle=0]{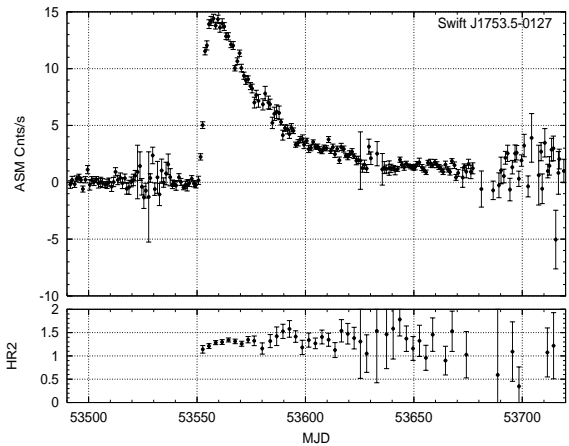}
\caption{ASM light curve (top panel) and the hardness ratio (HR2) (5-12 keV)/(3-5 keV) (bottom panel); hardness ratio is calculated from the 3-day averaged count rate in the two energy bands.}
\label{ASM-lc}
\end{figure}

The observations of the source by the All Sky Monitor of RXTE \citep{Levine} date from July, 1, 2005 to November, 28, 2005.
The light curve from the one-day averaged data of ASM is shown in figure \ref{ASM-lc}.
The profile of the light curve is a typical FRED (Fast Rise Exponential Decay).
The peak of the ASM light curve of the burst corresponds to 200 mCrab. 
The unabsorbed PCA flux in the 3-25 keV corresponding to the peak of the burst on July 6, 2005 (MJD: 53557) is 7.19e-09 ergs/s-cm$^{2}$, which corresponds to $L_{X}/L_{Edd} \simeq 0.05 (d/8.5~{\rm kpc})^{2}~ ({\rm M}/10~{\rm M}_{\odot})$.
The rise time of the light curve is found to be 8 days.
An exponential fit to the light curve gives an e-folding time of $\sim$30 days.
The e-folding time derived from the PCA light curve (31 days) for the energy range 3-25 keV matches with that of the ASM light curve and the light curve is found to deviate from the exponential decay after $\sim$50 days from the peak of the outburst .
The bottom panel of figure \ref{ASM-lc} shows the hardness ratio, the ratio between the count rates in the energy bands (5-12) keV and (3-5) keV.
The hardness ratio is calculated from the three-day averaged count rate in the two different bands so that the evolution of it can be seen clearly.
This hardness ratio is $\sim$1.0 at the start of the burst, increases to $\sim$1.5 and remains $>$ 1.0 throughout the burst, which indicates that the source has not entered the high/soft state throughout the burst.

\subsection{PCA and HEXTE data: Spectral analysis}
\label{sec:spectra}
We have extracted the PCA energy spectra from the standard-2 data which has an intrinsic resolution of 16s.
We have used the PCU2 detector data for the spectral analysis.
A systematic error of 1\% is accounted for in the data.
The spectra are extracted using FTOOLS V6.0.
The background is estimated using pcabackest V4.0 and subtracted from the data.
PCA response matrices are generated using pcarsp V10.1.
We have extracted the HEXTE spectra from one of the clusters of the instrument, the Cluster A.
We have used the standard mode data for which the spectral bins are in 64 channels with 16 seconds time bin.
For the fits to the combined data in the energy range 3-180 keV, we have considered the data from PCA in the energy range 3-20 keV and that from HEXTE in the range 20-180 keV.
In order to account for the uncertainties in the relative calibration of the PCA and HEXTE instruments, the normalization factor for the PCA data is frozen to 1 and that of the HEXTE data is allowed free for the fits to the combined data from both these instruments.
The n$_{H}$ value of the column density for interstellar absorption is fixed at 2.3 $\times$ ${10^{21}}$ atoms-cm$^{-2}$ \citep{Millera} for all the fits.

\begin{figure}
\includegraphics[height=8cm,width=6cm,angle=-90]{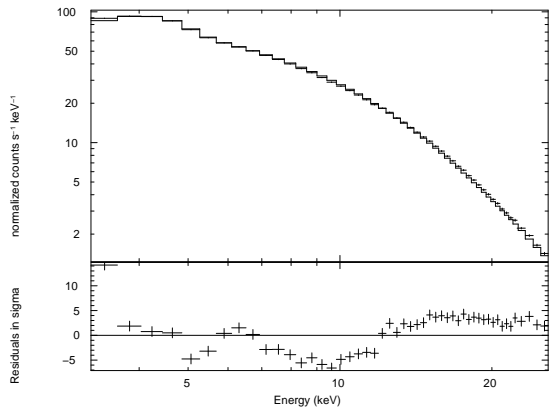} 
\caption{The spectrum from PCA data for one observation (obsid: 91423-01-01-04)fitted with a simple power law component is shown here. The residuals are shown in the bottom panel, in units of sigma.}
\label{pca-spec-nodisk-nosmedge} 
\end{figure}

First, we have used the energy spectra from the PCA data in the energy range 3-25 keV, for the fits with a simple power law.
One such fit for the observation during the peak of the burst is shown in the top panel of figure \ref{pca-spec-nodisk-nosmedge}, for which the photon index is found to be 1.83 as shown in table \ref{parameters}.
However, the fits to the data with just a simple power law show large residuals at lower energies.
This can be seen in the bottom panel of figure \ref{pca-spec-nodisk-nosmedge}.
To account for the residuals at the lower energies we have added a thermal component in the model.
We have tried different models for it, like the multicolor disk blackbody (diskbb in Xspec), the blackbody component (bbody in Xspec) as well as the blackbody component corresponding to the boundary layer emission of a neutron star (bbodyrad in Xspec). 
The fits with these different models for the thermal component do not show substantial difference.
We decided to fit the thermal component with a diskbb, because the radius of the surface of emission at the peak of the burst as estimated using bbodyrad is  $\sim$100 km for a distance of 8 kpc, which is unrealistic for a neutron star.    
After the inclusion of the thermal component, substantial residuals are still present around 7 keV. 
This is an indication of the presence of an Fe emission line along with an absorption feature.
We tried including the absorption edge component and a gaussian to the model, which removes the line feature but not the absorption feature completely. 
However, on the addition of the smedge component ("smedge" in Xspec \citep{Ebisawa}, which accounts for the absorption feature), the emission feature also disappears and the fits are found to improve substantially.
From the reduced chi-square value and the residuals at the high energy end, we find that there is a need to fit the data with a high energy cutoff component to get a better fit.

\begin{figure}
\includegraphics[height=8cm,width=6cm,angle=-90]{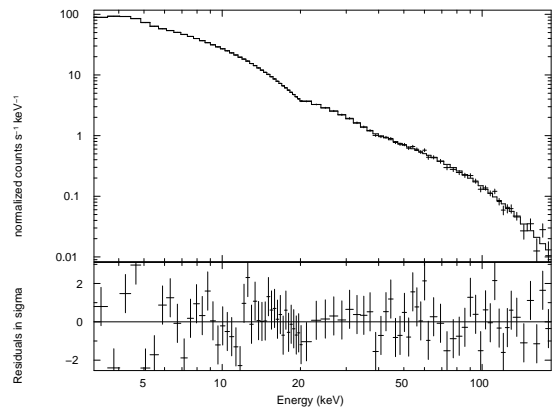}
\caption{The spectrum from PCA and HEXTE data, for the observation (obsid: 91423-01-01-04) shown in figure \ref{pca-spec-nodisk-nosmedge}, fitted with a combined model of simple power law, a smeared edge (smedge), high energy cutoff and diskbb components is shown here. The residuals in the bottom panel are in units of sigma.}
\label{pca-hexte-spec} 
\end{figure}

We have then combined the PCA and the HEXTE data for the final fits.
The spectral model that we have used is a combination of a simple power law, a smeared edge (smedge), a high energy cutoff, a diskbb component and a photon absorption parameter to account for the interstellar absorption.
The spectral fit to the combined data from the PCA and the HEXTE data with this model is shown in figure \ref{pca-hexte-spec}.  
For the observation at the peak of the burst, the power law spectrum has an index of 1.76{\tiny ${^{+0.01}_{-0.02}}$}, the temperature of the disk (T$_{in}$) is found to be 0.42{\tiny ${^{+0.09}_{-0.06}}$} keV and the high energy cutoff is estimated to be 38.2{\tiny${^{+10}_{-8}}$} keV.
The absorption edge derived from the smedge component is found to be at $\sim$7 keV and this does not vary throughout the burst.
The power law component is the dominating factor in the spectra. 
The thermal component though necessary for the fit at low energies, contributes less than 2\% to the total flux in the 3-25 keV range.  
The inclusion of thermal component and the high energy cutoff are required for the first 35 days of the observations, after which only the power law component and the smedge component are required. 
Eighty days after the peak of the burst, the smedge component also becomes insignificant. 
The data sets after this are therfore fitted with a simple power law and the absorption parameter. 

In order to explain the underlying physics, we have fitted the data with the comptonization model (compTT in Xspec) \citep{Titarchuk}, which describes the comptonization of the soft seed photons from the disk by the hot plasma.
Though the uncertainties on the parameters are large, the fits to the complete set of observations indicate an average temperature of the hot corona to be $\sim$ 40 keV and the optical depth $\sim$ 1.4.
Inspite of the caveats of the large errors, the average value of the temperature of the seed photons from the accretion disk derived from this model, which is $\sim$0.4 keV, matches the temperature of the accretion disk derived from the fits mentioned above.
The fit parameters for the different models are given in table 1.

\begin{table*}
 \centering
 \begin{minipage}{190mm}
  \caption{Table showing the fit parameters and the reduced chi-square for the fits to the PCA and HEXTE data for the observation (obsid: 91423-01-01-04) during the peak of the burst}
\label{parameters}
  \begin{tabular}{lccccccc}
  \hline
\multicolumn {8}{c}{Fits for PCA (3-25 keV) data} \\
\hline
Model	&PI&	T$_{in}$&E$_{cut}$ & T$_{0}$ & KT$_{e}$	& $\tau$ & reduced-	\\
        &  &	(keV)	&(keV)	 	&	(keV)	& (keV) & & chi-square\\
\hline
PL	&  1.831 & -	&-&-&-&-& 15.38\\
PL+diskBB&  1.744 & 1.096 &  -&-&	-&-&  5.68\\
PL*Smedge+diskBB&  1.78\tiny{${^{+0.006}_{-0.006}}$} & 0.38\tiny{${^{+0.079}_{-0.057}}$}&-&-&-&-&  1.67\\
\hline
\multicolumn{8}{c} {Fits for PCA+HEXTE (3-180 keV) data} \\
\hline
PL*Smedge*HEcut+diskBB	&  1.76\tiny{${^{+0.012}_{-0.015}}$} &  0.42\tiny{${^{+0.086}_{-0.058}}$} & 38.2\tiny{${^{+9.8}_{-7.7}}$} &-&-&-& 1.3\\
compTT*smedge+diskBB	&	-	& 0.43$\pm$0.04	& -	& 0.6$\pm$0.13	&  44.6$\pm$8.7 & 1.08$\pm$0.23 & 1.33\\
\hline
\multicolumn {8}{l}{Explanations for the abbreviations in the table are as follows:}
\end{tabular}
\\
PL - Power Law; HEcut - High Energy cutoff; PI - Photon Index of the power law; T$_{in}$ - Temperature of the disk; T$_{0}$ - Temperature of the disk which is the source of the seed photons; KT$_{e}$ - Temperature of the plasma, ie. the corona; $\tau$ - optical depth of the plasma\\
\end{minipage}
\end{table*}

\subsection{PCA data: Timing analysis}
\begin{figure}
\includegraphics[height=7cm,width=5cm,angle=-90]{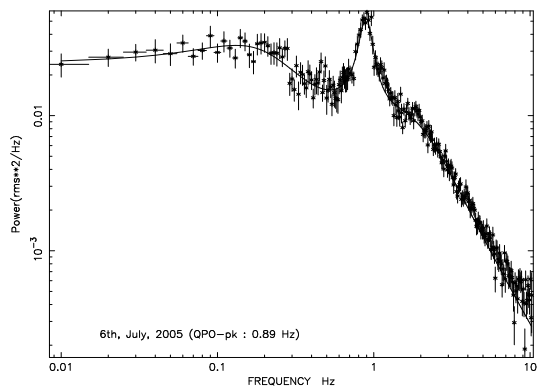}
\caption{Power density spectrum showing the low frequency QPO, for the observation (obsid: 91423-01-01-04) at the peak of the burst.}
\label{qpo} 
\end{figure}

Timing analysis is done using the event mode PCA data which has a time resolution of 125 $\mu$s.
Power Density Spectrum (PDS) is generated for each observation using ftools.
The power spectra are normalised such that their integral gives the squared rms fractional variability ((rms)**2/Hz), with the expected white noise level subtracted.
The PDS thus obtained are fitted with a sum of Lorentzians.  
The power density spectrum obtained from the PCA data during the peak of the burst is shown in figure \ref{qpo}.
A prominent low frequency QPO (LFQPO) at 0.891$\pm$0.008  Hz. with a Q-factor of 4, and an amplitude of $\sim$0.06 is observed.
There are no kHz QPOs found in the PDS.
The peaked noise and the band-limited noise are fitted with other Lorentzians.
The total rms power of the PDS is found be 23\% at the peak of the outburst.
In order to determine the energy dependence of the LFQPO, we generated the PDS for different enegry bands.
We find that the QPO is prominent in all the data sets in the energy range of 3-15 keV.
Its presence at higher energies is evident only in few data sets.
\subsection{Evolution of parameters}
\begin{figure}
\includegraphics[height=14cm,width=8cm,angle=0]{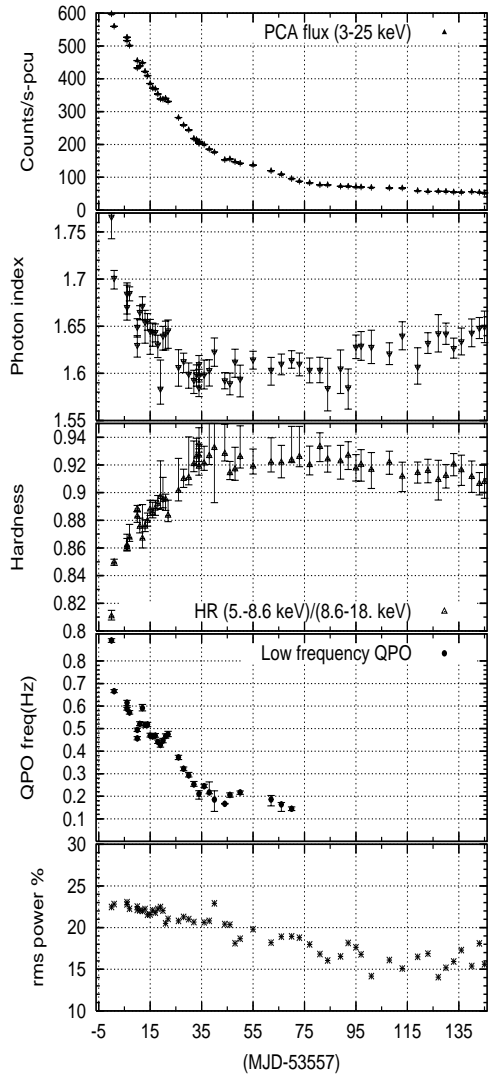}
\caption{The time evolution of PCA count rate, photon index, hardness ratio, QPO and the rms power are shown from the top to the bottom panel respectively.}
\label{5-para} 
\end{figure}

We have ploted different parameters and the observed flux as a function of MJD in figure \ref{5-para}.
This figure shows the time evolution of various parameters like the PCA count rate (in the energy range 3-25 keV), the photon index, the hardness ratio (the ratio of the PCA count rate in the energy bands (8.6 to 18.0) keV to (5. to 8.6) keV, as defined by \citet{Muno,Remillard}), the low frequency QPO and the rms amplitude of the power density spectrum in the frequency range 0.1 to 10 Hz.
It may be noted that the entire set of data available is from the peak to the decay of the burst.  
    
The power law component is dominant throughout the decay of the burst.  
The photon index is found to be 1.76 at the peak of the burst and decreases to 1.70 within a day although the flux decrease observed is within the level of 1$\sigma$.  
After this abrupt decrease the  photon index of the power law further decreases almost linearly from 1.7 to 1.6 within a span of 30-35 days.   
Later the photon index hovers around 1.61 till about 90 days after which there is a slow trend of increase seen towards the end of the observations.
The hardness ratio is also found to indicate a similar behaviour, 
with HR showing an abrupt increase from 0.81 to 0.85 within a day followed by an almost linear increase from 0.85 to 0.92 within 30-35 days and then leveling off at 0.92 till 90 days.  
The  tendency for HR to decrease gradually after 90 days upto the end of the burst is also similar to the photon index.
The spectral softening observed after 90 days is very gradual with time and continues till the end of the observations.
This behavior of spectral softening is observed in few other sources like XTE J1550-564 \citep{Sturner,Belloni}.

The centroid frequency of the QPO is found to be 0.891 Hz at the peak of the outburst and is found to decrease abruptly to 0.66 Hz within a day and then it decreases almost linearly from 0.66 Hz to 0.2 Hz within a span of about 35 days.   
Later, the QPO is visible only intermittently at lower values of about 0.2 Hz and is observed only till 70 days, after which it is not detectable.
The rms power of the PDS (0.1 to 10.0 Hz) however exhibits a more uniform trend with a slow decrease from 23 to 15\% till the end of the burst.  
This trend is also seen if the frequency range of the PDS is extended down to 0.01 Hz.

\begin{figure}
\includegraphics[height=12cm,width=8cm,angle=0]{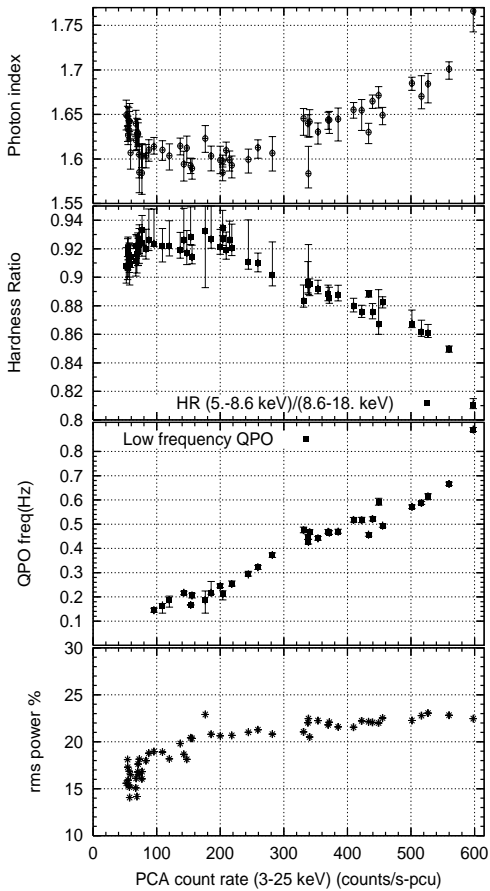}
\caption{The photon index is shown at the top panel, the hardness ratio (8.6-5.0 keV)/(8.6-18.0 keV) derived from the PCA data in the second, the QPO frequency in the third and the rms power at the bottom.}
\label{4-para} 
\end{figure}

In order to study the variation of these parameters as a function of flux,  we plot the same in figure \ref{4-para}.
It may be noted that since the observations cover the decay portion of the burst, the time line in this figure goes from right to left.
The softening of the spectra at low flux levels (towards the end of the burst) can be clearly seen in the top two panels of figure \ref{4-para}, which shows the photon index and the hardness ratio as a function of PCA flux.
The QPO frequency follows a linear trend as a function of flux, which is seen in the third panel.
The rms power of the PDS shows an abrupt decrease at low flux levels as shown in the bottom panel of the figure.
The time evolution of the disk parameters like the temperature of the disk, the inner radius of the disk etc. are difficult to comment on as the uncertainities on these parameters are large.
The high energy cutoff is found to be about 40 keV on an average and a straight line fit to the values during the entire outburst does not show any systematic change during the decay.

\section{Discussions}
\subsection{On the nature of the source}
The fits to the XMM-Newton data by \citet{Millera}, give an n$_{H}$ of $\sim$ 2.3 X ${10^{21}}$ atoms/${cm^{2}}$, which is a nominal value for galactic interstellar absorption.
The optical observations \citep{Halpern} of the source reveal the presence of a bright optical counterpart with R$\sim$15.8 magnitude indicating the system to be a low mass x-ray binary (LMXB), with a reddened companion star.
Spectral analysis of the source indicating the hard spectrum with a power law dominance and the timing analysis of the data indicating the presence of low frequency QPOs at $<$ 1 Hz are typical features observed in BHCs.
There are no kHz QPOs found in the power spectrum.
Figure \ref{qpo} shows the low frequency QPO at $\sim$0.891 Hz.
All the above points indicate that this source could be a galactic LMXB with a stellar mass black hole as the compact object more likely than a neutron star.

In addition, the fits to the thermal component in the spectra with a black body component ("bbodyrad" in Xspec) for a normalized surface area, corresponding to the boundary layer of a neutron star, gives an estimate of the radius of the compact object $\sim$100 km, for a typical distance of 8 kpc, at the peak of the burst and is found to be $>$20 km till the end of the outburst.
From this, it can be surmised that the probability for the compact object being a neutron star is almost negligible.

\subsection{On the spectra and the underlying physics of the burst}
The low/hard spectral state of  black hole candidates (BHC) is typified with a dominant power law component, contributing $>$ 80\% of the total flux in 2-20 keV range, having a spectral index between 1.4 and 2.1.
The total rms power (r), in the power density spectrum, integrated over the frequency range 0.1-10 Hz is strong with r $>$ 0.1 \citep{Remillard}.

We find that for SWIFT J1753.5-0127, the power law component dominates the spectra throughout the burst with the spectral index decreasing from 1.76 to 1.6. 
Further, the thermal component, though present during the bright phases of the outburst, contributes only $<$ 2\% of the total flux in the 3-25 keV energy range.  
This, along with the presence of a low frequency QPOs $<$1 Hz and the rms power in the frequency range 0.1-10 Hz being $>$ 10\% (shown in figure \ref{5-para}), clearly indicates that the source is locked in the low/hard state throughout the outburst and never made it to the high/soft state.

In all the LHXT outbursts observed till date, the presence of a hot accretion disk component in spectra of the 3-25 keV energy range is not reported.
We find a diskbb component at 0.4 keV and in addition the presence of a smedge component, which is usually attributed to be the disk reflection component. 
The inner radius of the disk (estimated from the normalization factor of diskbb component) turns out to be $\le$4R$_{g}$ for a black hole of mass $10~{\rm M}_{\odot}$.
We find that the contribution of the disk component is seen only till day 35 and the smedge component is seen till 80 days after the peak of the outburst.
We suppose that the reason for the lack of a thermal component in the later part of the burst in the 3-25 keV range is due to the reduced sensitivity of the RXTE-PCA instrument at energies below 3 keV and that, the accretion disk continues to exist even towards the end of the burst.
The presence of the accretion disk during the low/hard state approaching quiescence for this source, SWIFT J1753.5-0127, was strongly suggested by \citet{Millera}, with simultaneous RXTE and XMM-Newton observations, after about 118 days from the last observation of the outburst discussed in this paper. 
The fits to the XMM-Newton data by \citet{Millera} imply the presence of an accretion disk extending near to the inner stable circular orbit with an R$_{in}$ $\le$ 6R$_{g}$ for a black hole of mass $10~{\rm M}_{\odot}$. 
All these provide the evidence for the presence of the inner accretion disk in a low/hard state x-ray outburst.

Fits to all the RXTE data sets of the outburst of the source, SWIFT J1753.5-0127, show that the spectrum is however, dominated by the power law component, which can be described by the inverse Comptonization of seed photons from the accretion disk by a hot plasma near the central compact object.
This picture is supported by modeling the data using the comptonization model ("compTT" in Xspec), the results of which show an accretion disk of temperature $\sim$0.4 keV and a hot corona of temperature $\sim$40 keV with an optical depth $\sim$1.4. 
This Comptonisation region could be  a corona, a Compton cloud or a post shock region as referred in the literature.  

\begin{figure}
\includegraphics[height=8cm,width=6cm,angle=-90]{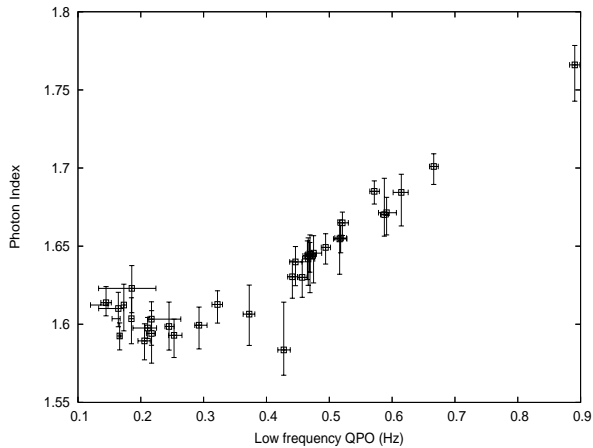}
\caption{Photon index as a function of low frequency QPO}
\label{pi-vs-qpo} 
\end{figure}

Low frequency QPOs less than 1 Hz are observed upto about 35 days. 
The QPOs seen in the PDS of all the observations are found to be from photons predominantly in the energy range 3-15 keV.
Since there is a contribution from photons at high energies, it appears that the QPOs have both a thermal and non-thermal component.
Simple Keplerian inflow of matter predicts higher frequencies for a $10~{\rm M}_{\odot}$ black hole.
The centroid frequency decreases with time and appears to be linearly correlated with the flux (figure \ref{4-para}).
This correlation does not favour the global normal disk oscillation model \citep{TO} for the origin of QPOs in this source, as also indicated by \citet{Zhang}. 
The trend of photon index and QPO correlation as seen in figure \ref{pi-vs-qpo} is similar to that predicted by the transition layer model of \citet{TF}. 
The optical depth ($\tau$) of the transition layer, is estimated to be $\sim$1.8 from the figure 6 in \citet{TF} for a photon index of 1.7.
This is comparable to the value of 1.4 for the optical depth of the corona derived from the CompTT model as applied to these observations.
Using figure 4 in \citet{TF}, which relates LFQPO frequency to the outer radius of the transition layer, the 0.9 to 0.7 QPO frequency decrease observed in our data corresponds to 20 to 30 R${_s}$, which translates to 600-750 km for a $10~{\rm M}_{\odot}$ black hole.
This size of TL is however much larger than the estimated R$_{in}$ from the diskbb model which is of the order of 100 km.
Therefore, the current observations can match the size of the transition layer only if the QPO-vs-TL size relationship can be proportionally reduced for a lower mass black hole, of the order of $3~{\rm M}_{\odot}$ black hole.  
However, we hasten to add that although the range of spectral index predicted by the transition layer theory for low/hard state ($\Gamma$ = 1.6 $\pm$ 0.2) applies to our observations, the observed dependence of spectral index on flux is not predicted by the model.
In addition, this model predicts that the the low frequency and high frequency QPOs are correlated.  
We do not observe any high frequency QPO for this source. 

\subsection{On the light curve of the burst}
The profile of the light curve of this source, SWIFT J1753.5-0127, is of a FRED (Fast Rise Exponential Decay) type.
The e-folding time of the light curve is $\sim$31 days which is a nominal value found in x-ray novae outbursts.
FRED type light curves of the x-ray outbursts are supposed to be originating due to the instabilities in accretion disk \citep{Chen}.
This can be explained either by the Disk Instability Model (DIM) \citep{Lasota} or due to the disk diffusion propogation model \citep{Wood}.
The correlation of QPO frequency with source flux as seen in figure \ref{4-para} does not favour the disk diffusion propogation model \citep{Wood}.

Since the profile of the light curve is a FRED, which is associated with the disk, it can be considered that the outburst could have been triggered due to some instability in the disk.
It may be noted that the e-folding decay time of this outburst (31 days) is also associated with similar timescales on which several spectral parameters show a change in behavior indicating a strong correlation between the spectral and light curve characteristics. The disk probably starts receding with the abrupt decrease in the spectral index within a day, and by the end of 35 days it cools to the extent that it is not detectable in the 3-25 keV region.  
This change is indicated by the increase in hardness and decrease in the QPO frequency within a day of the outburst.

A similar FRED profile has been seen in the 40-150 keV light curve of GRO J0422+32 during its 1992 low/hard state x-ray outburst detected by BATSE on CGRO (Burst And Transient Source Experiment on Compton Gamma Ray Observatory) \citep{Van}.
Since there was no soft x-ray coverage during the early stages of the burst, it could not be proved that an ultrasoft disk component existed or not \citep{Pietsch}.

The light curves of most of the x-ray outbursts in the low/hard state of the sources observed in the 2-10 keV energy range are found to be of triangular shape with or without a plateau.
None of these outbursts are found to be reported with an ultrasoft component associated with an accretion disk in their spectra.
However, the presence of a cool outer disk can be found implicitly, from the reflection components in the spectra.
Whether a common instability model with difference in parameters can explain both the FRED profile and the triangular profile for low/hard X-ray outbursts is still an open question.

\subsection{On the possible explanations for a low/hard state x-ray outburst}
While \citet{Brocksopp} try to look for a possibility of explaining the behavior of these LHXTs as part of the outburst mechanisms associated with other canonical SXTs, \citet{Meyer} gives an explanation in terms of short orbital periods, which result in less mass accumulation in a smaller accretion disk giving rise to relatively low peak luminosities during the outbursts.

It is stated \citep{Meyer} that for low peak luminosities, ie. low mass flow rate in the disk, coronal evaporation truncates the thin disk even in the outburst and an advection-dominated accretion flow (ADAF) occurs resulting in a spectrum that remains hard.
Also the evaporation efficiency is stated to be proportional to the mass of the compact object and larger the mass, larger is the truncation radius of the inner disk \citep{Meyer}. 
The behavior of SWIFT J1753.5-0127 can also be explained by the ADAF model, if the compact object is a less massive black hole, resulting in lesser evaporation efficiency allowing the disk to extend near the inner stable circular orbit accounting for the thermal component, while having a contribution of evaporation of the disk to give rise to the overwhelming hard state of the outburst.
We also see a softening towards the end of the burst which is seen in figure \ref{5-para} and more clearly seen in figure \ref{4-para}.
Spectral softening at low luminosities is predicted by ADAF model \citep{Esin}.
This could imply that it is possibly ADAF which is being observed at the low flux levels of the burst approaching quiescence.

There are some sources which have been observed only in low/hard state outbursts.
They are GRO J0422+32 \citep{Van}, GRS 1716-249 \citep{Hjellming}, GS 2023+338, 1E1740.7-2942 and GRS 1758-258, amongst which the last two have been reported to be persistent sources \citep{Grebenev, Cui}.
The reason for a source to show low/hard state x-ray outbursts only and not having any canonical outbursts could be something to do with the orbital parameters of the system as such.
Such systems can be classified as Low/Hard state X-ray Transients (LHXTs), as mentioned by \citet{Brocksopp}. 
There are also sources, which have had state transitions in canonical outbursts, at times showing low/hard state outbursts.
A few of them are XTE J1118+480, XTE J1550-564, GX 339-4, Cyg X-1 etc.
The low/hard state x-ray outbursts shown by these sources can be classified as "failed outbursts" in the canonical SXTs, which could be due to low mass accretion rate or due to discrete accretion events.

While there are many models to explain the varied outbursts from different sources, whether a common model will be able to explain all these different natures of the outbursts or not, is still an open question.

\subsection{Conclusions}
Spectral and timing analysis of RXTE observations of the source, SWIFT J1753.5-0127, show that the source was in the low/hard state throughout the outburst.
The physical picture of the burst is explained by the comptonization of the seed photons by the hot corona near the compact object.
The FRED profile of the light curve, usually uncommon in low/hard x-ray outbursts and the presence of a multicolor black body component in the fits to the spectra imply a likely association of the burst with the instabilities in the accretion disk.
The low frequency QPOs $<$1 Hz and the rms power between 10 and 30 \%, along with the spectral behavior of the source with an optical counterpart of magnitude R$\sim$15.8 indicates this source to be a LMXB with a stellar mass black hole, most likely, as the compact object.
Thereby, this source falls in the category of SXTs that show low/hard state x-ray outbursts.

The QPO frequency observed during the first 35 days after the peak of outburst is almost linearly correlated with the flux and the photon index.  
During the same period a spectral hardening and diminishing contribution of disk is also found.  
These factors rule out some of the current models for outbursts.  
There is also a softening of spectrum seen towards the end of the burst, accompanied with very small change in source flux.  
While the transition layer model and the ADAF model can explain some of the observed features, both the models indicate that the compact object in this system is most  probably a low mass black hole.
\section{Acknowledgements}
We thank Dr. P. Sreekumar for discussions and overall guidance.

This research has made use of the data obtained through High Energy Astrophysics Science Archive Research Center online service, provided by NASA/Goddard Space Flight Center.

\label{lastpage}

\end{document}